\documentstyle[epsfig]{article}
\title{Phase states in amorphous carbon\footnote{to be published in
{\it Nanostructured Carbon for Advanced Application,} 185-198, ed. G. Benedek
et al. (2001 Kluwer Academic Publisher, Netherlands).}}
\author{A.S.Bakai, Yu.A.Turkin and M.P.Fateev\\
\\ {\it NSC Kharkov Institute of Physics and Technology, Akademichna 1,
61108 Kharkov, Ukraine}
}

\begin{document}
\maketitle
\begin{abstract}
Amorphous carbon (a-C) formed under energetic atom deposition and by
cooling of a melt by MD simulations in a wide $P,T$-range is
investigated.  Simulations  of a-C formation with atomic beam
deposition reveal a sharp GLC-to-DLC transition with $z=3.35 $.
The average coordination number,  $ z $, depends on
the depositing atom energy.  Voronoi polyhedra of  the
3-coordinated atoms changes dramatically in the vicinity of the
transition while for the 4-coordinated atoms they are changing a
few. The elastic moduli along with densities of states of
electrons and phonons in the GLC and DLC states are investigated.
Changes of a-C with temperature and pressure evolution and GLC-DLC
transformations are studied. The map of isoconfigurational states on
the ($P,T$)-plane shows regions of GLC and DLC phases stability.
\end{abstract}

\newpage
\section{Introduction}
Amorphous carbon (a-C) in solid state is a rather big family of carbon
materials with many types of short-range and medium-range orders
(SRO and MRO). From thermodynamic point of view any a-C is a non-ergodic
and non-equilibrium system. The diffusional structure relaxation time in
graphite and diamond at $ T<10^{3}$ K  is
huge because the activation energy of this process is estimated
to be larger than $ 7$ eV.  In a-C the structure relaxation
processes are also very slow at those temperatures. Therefore
many forms of a-C are rather stable to be used and investigated.
Structure of solid a-C is a continuous random network (CRN). The
local order (LO) of an atom is determined by configuration of its
covalently bonded nearest neighbors. Because of polyvalence the
a-C network possesses an alternating LO. The topologic disorder
of the network originates from both alternating LO and random
local distortions. The MRO of CRN is mainly determinated by
correlations of atoms of the same LO and by orientational
correlation of the Voronoi polyhedra. From general point of view
there are no restrictions on the ratios $ c_{2}/c_{3} $ and
$ c_{3}/c_{4} $ ($ c_{2},c_{3},c_{4} $ are
concentrations of $sp$-, $sp^{2}$-, $sp^{3}$- fractions), as well as
on the range and characteristics of MRO.  The structure
properties of a-C are determined by kinetics of its formation.
For this reason different forms of a-C are characterized by the
initial (essentially non-equilibrium) state of C and by the
method of its preparation. Here we are concentrated on
investigations of topology and properties of a-C obtained by atom
(ion) beam deposition deposition and by modelling of cooling of a
melt  in a wide $ P,T $-range. These forms of a-C have a lot of
applications [1-4]. Experimental information
concerning the microscopic structure of these materials is
incomplete, but big variety of their macroscopic properties is
observed.  In \cite{ro} the a-C films were separated on the
graphite-like (GLC) and diamond-like (DLC) rather qualitatively:
those films which have macroscopic properties (density, hardness,
electric and heat conductivity) similar to that of diamond are
treated as DLC.  Otherwise the films are GLC. It was believed
\cite{r} that mainly the big concentration of $ sp^3$ bonding
is responsible for the diamond-like properties. Meanwhile
computer simulations \cite{k} showed up that the distorted
$sp^2$-network with low $sp^3$  fraction
can have diamond-like macroscopic properties.  It was still
unclear whether there are topological and thermodynamic criteria
of the GLC and DLC state classification. The canonical
thermodynamic approach is not valid in the case of a non-ergodic
system \cite{p}.  Therefore we start from the polyamorphizm
problem formulation.  Then molecular dynamic (MD) computer
simulations are used to get an answer whether the polyamorphous
states of a-C (GLC and DLC) can be identified or not. It turns
out that the positive answer does exist at least in the case of
a-C with Tersoff's or Brenner's classic empiric potential. The
classical empiric potential simulations (CEPS) of a-C formation
with atomic beam deposition reveal that a sharp GLC-to-DLC
transition takes place with $z=z^{*}=3.32 $, $z $
is the average coordination number of a-C. It depends on the
depositing atom energy, $E_a $. We have compared the densities
of states (DOS) of electrons and phonons in the GLC and DLC
states.  To calculate them the tight-binding MD (TBMD)
simulations were performed.

Further changes of
a-C with temperature and pressure evolution and GLC-to-DLC
transformations in the CEPS are investigated. The map of
isoconfigurational states on the ( $ P,T  $)-plane
shows regions of GLC and DLC phase stability. Specific volume and
potential energy of a-C as functions of  $ P,T $
along with the map of isoconfigurational states allow to
determine glass-to-liquid transition temperature,  $  T_g $, vs
$P $. In result the phase diagram of a-C is constructed.  It has
to be pointed out from very beginning that all of the obtained
results concern the model a-C with the chosen classical empiric
potential or with the TB interactions. Nevertheless we believe
that they mimic reality at least qualitatively.

\section {Polyamorphizm of a-C}
The term ``polyamorphizm" was introduced in analogy with
``polymorphizm" in \cite{pn} and then was reintroduced
independently by others \cite{a}.  Literally it means that a
system can exist in two different amorphous phase states of the
same composition. Concerning solid amorphous states the given
definition became invalid because many non-ergodic structure
states with different structure and thermodynamic properties can
be stable at the same $ (P,T) $-point for much longer
than the observation time. Moreover the number of the possible
structure states is exponentially large (see below). Therefore
the free energy has a lot of comparable minima in the phase space
(see e.g. \cite{a,an}). The minima set can be ordered by the
following way. A subset of the minima of a comparable depth which
are separated by low barriers and correspond to states of
different SRO's forms a basin of the structure states. The
basins are separated by higher barriers. Those basins which have
comparable depths and similar structure properties of states
within them do form megabasins \cite{a}. The number of
structure states within of a megabasin is determined by
configurational entropy or complexity \cite{p},  $\zeta_{i}$
($i$ is the index numbering the megabasins):
\begin{equation}\label{aa}
W(N)=\exp(\zeta_{i}N)+W_{ne}(N)
\end{equation}
Here $ N$ is the number of atoms and $W_{ne} $ is the number
of the structure states which do not contribute into the
configurational entropy in thermodynamic limit,
\begin{equation}\label{bb}
\lim_{N \to \infty}\ln W_{ne}/N=0
\end{equation}
Transitions within a basin are governed by the short range atomic
rearrangements. The characteristic time of these rearrangements,
$\tau_{SRO} $, is the shortest of the structure
transformation times.  MRO transformation needs correlated
structure rearrangements which can be connected with formation of
stressed regions or inner boundaries.  This process needs also
the short range diffusion. The characteristic time of this
transformation, $ \tau_{MRO}$, as a rule,
is much longer as compare with  $ \tau_{SRO}$.
The long-range ordering time,  $  \tau_{LRO} $,
is the largest of the structure relaxation times. The
observation time, $t_{obs}$, has to be much
shorter than  $ \tau_{LRO} $, otherwise a stable crystalline
phase forms. The averaged within $i$-th megabasin free energy,
$<G_{i}> $, depends on $P,T $. It is the thermodynamic
quantity which characterized the amorphous phase and makes
understandable the polyamorphism thermodynamics. The difference
\begin{equation}\label{cc}
\Delta G_{ij}=<G_{i}>-<G_{j}>
\end{equation}
determines stability of the amorphous phases.
Crossover of the ``coexisting curve",
\begin{equation}
\Delta G_{ij}(T,P)=0,  \nonumber
\end{equation}
is the necessary condition of the polyamorphic transformation.
Let us denote by  $T_{12}(P)$ the
coexistance temperature of phase states $1$ and $2$.  The
transformation is observable with
\begin{equation}\label{dd}
\tau_{LRO}>>t_{obs}>>\tau_{SRO}^{12}
\end{equation}
or
\begin{equation}\label{ii}
\tau_{LRO}>>t_{obs}>>\tau_{MRO}^{12}
\end{equation}
Here $\tau_{SRO}^{12},\tau_{MRO}^{12}$,
are times of transition from a state within the megabasin $1$
into a state within the megabasin $2$.  Because of the
multiplicity of the structure states within the basins,
(\ref{aa}), one can expect that the polyamorphic
transformation in result of a sequence of the short range
rearrangements is possible and the condition (\ref{dd}) is
satisfiable.  In the case of a-C it seems that this condition can
be fulfilled at least for polyamorphic states which differ mainly
by LO. MD simulations mimic the kinetics of the structure
formation evolution. The time of  state$1$ to state$2$
transformation which controlled by the short-range rearrangements
can be presented in the following form
\begin{equation}\label{ff}
\tau_{SRO}^{12}=\tau_{SRO}^{0}\sinh (\Delta
G_{12}\beta)=\nu_{0}^{-1}\exp(-h_{SRO}\beta)\sinh (\Delta
G_{12}\beta)
\end{equation}
where  $\beta \equiv 1/k_{B}T $,  $k_B $
is the Boltzmann's constant;  $\nu_0 $ is
the frequency of ``attempts" (it is about the Debye frequency);
$h_{SRO}$ is the activation enthalpy of the short-range
rearrangements. $\Delta G_{12} $, the
difference of the free energies of the states, is the
thermodynamic driving force of the transformation. This
expression is valid with $\Delta G_{12}\beta <<1$.
\begin{figure}
\begin{center}
\epsfig{figure=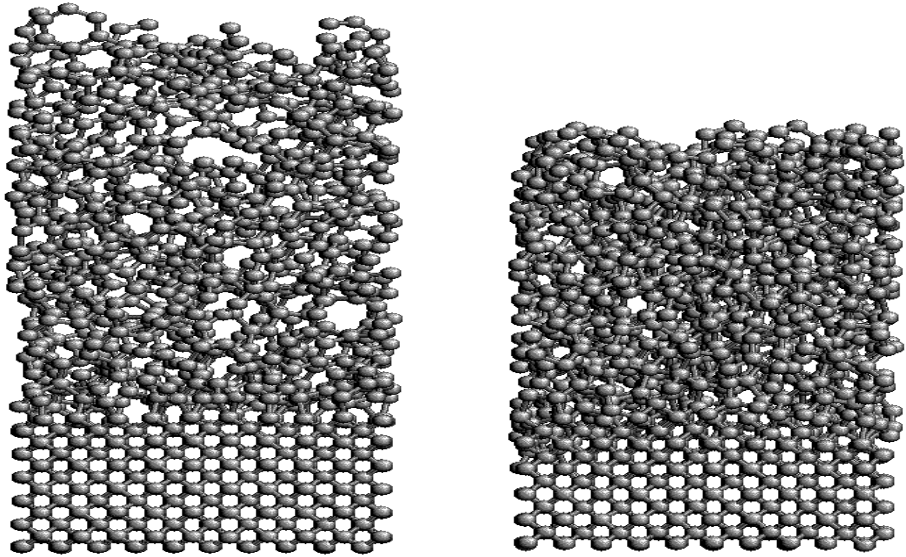,width=12cm}
\caption{\label{Fig1} a-C films deposited on diamond substrate from atom
beam with energy (a)$1$~eV, and (b) $40$~eV;\ $N=1800$}
\end{center}
\end{figure}
The~essential difference of the specific volumes,
compressibilities and thermal expansion coefficient of graphite
and diamond allows to expect that these quantities along with
$c_{2},c_{3},c_{4} $ are enough to identify GLC and DLC.

\section{Topologic transition in a-C deposited from atom beam}
We have performed the CEPS (with Tersoff's potential) of a-C
deposited from atom beam. The structure and properties of the
obtained ``samples" (with $N$ up to $1800$) were investigated.
Some results of similar MD simulations of Kaukonnen et al. (with
$N=516$) are reported in \cite{kn}. These results are
confirmed in our investigations but our main goal is the assumed
polyamorphism of a-C.  Beam deposition on diamond substrate at
$T=300$~K produces low-density and high-density samples (see
Fig. 1), dependently of  $E_a $.
Density and coordination number,  $z $, vs  $E_a $
are shown on Fig. 2.  Besides of the dramatic increasing of
the density at very low  $E_{a}<10$ eV, no specific
behavior of these quantities are observed.  Pair correlation
function of 3-fold and 4-fold coordinated atoms show that
independently of  $E_a $ correlations are significant only
up to second coordination shell and that average pair correlation
radius,  $r_c $, is about $2.5\AA$.
\begin{figure}
\begin{center}
\epsfig{figure=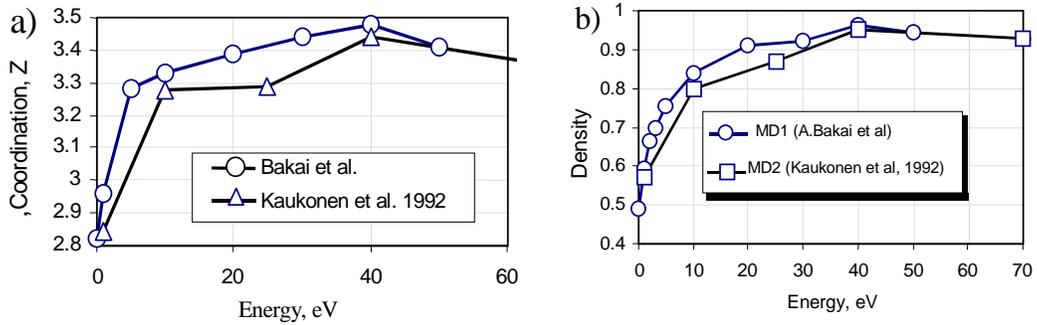,width=14cm}
\caption{\label{Fig2} (a)Coordination number,
and (b) density (in units of diamond density) as functions of
the energy $E_a$}
\end{center}
\end{figure}
Apparently, SRO determines
properties of this a-C. When the density, and  bulk, shear and Young moduli,
$\mu ,K $ and  $Y $  are plotted vs  $z $ (Fig. 3),
all these quantities ``jump" up to $ 30\%$ in the vicinity of
$ z=z^{*}=3.32$.  It looks like a phase transition on  $z $.
Because the pair correlation functions show up no dramatic
changes in the vicinity of  $ z^{*} $ it is clear that SRO
changes are responsible for the observed singularity.

Voronoi and Deloney divisions (they are dual each to other) are
useful in topology investigations. The results of analysis of
Voronoi polihedra are presented on Fig. 4 and Fig. 5. Voronoi polihedra of
4-fold coordinated atoms are almost unchanged in shape and volume
at any  $z $.  Voronoi cells of the 3-fold coordinated atoms
are distorted three-angled prisms at any  $z $ but their
volume changes essentially. The Voronoi polihedra of the 3-fold and 4-fold
coordinated atoms at $z>z^{*}$ are shown on Fig. 4.

\begin{figure}
\begin{center}
\epsfig{figure=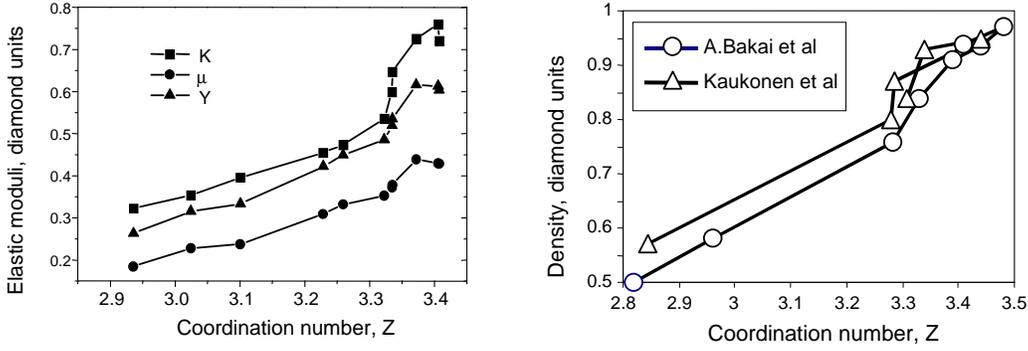,width=14cm}
\caption{\label{Fig3} (a) Elastic moduli, and (b) density as functions of the
coordination number}
\end{center}
\end{figure}
At low  $E_a $ the average volume of
cell of 4-fold coordinated atoms,  $v_{4}=6.92{\AA}^3 $, is
something larger than that with  $z>z^{*} $,
($v_{4}=5.74{\AA}^3$, due to contribution of micropores in the
former case.
\begin{figure}
\begin{center}
\epsfig{figure=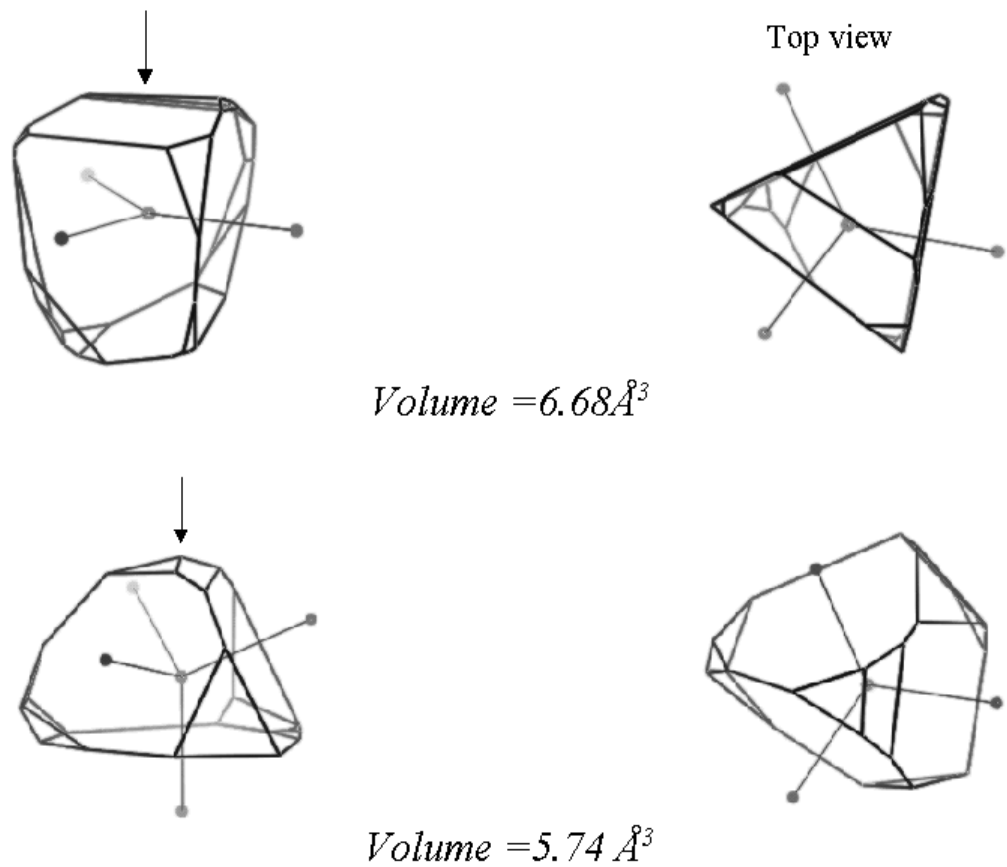,width=14cm}
\caption{\label{Fig4} Voronoi polyhedra of film deposited from $40$ eV
carbon beam}
\end{center}
\end{figure}
\begin{figure}
\begin{center}
\epsfig{figure=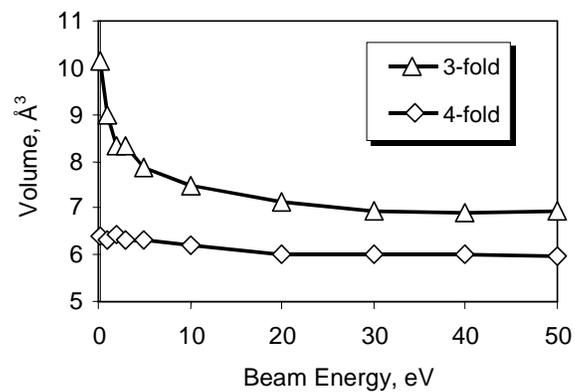,width=8cm}
\caption{\label{Fig5} Volumes of Voronoi polyhedra of the
3-fold and 4-fold coordinated atoms vs the beam energy, $ E_a$}
\end{center}
\end{figure}
Note that $ v_4=5.63{\AA}^3 $ for diamond at  $T\rightarrow 0 $.
The specific volume of 3-fold coordinated atoms dramatically
depends on  $E_a $. It is equal to $9.26{\AA}^3$ at  $z<z^{*} $
and to $6.68{\AA}^3$ with $z>z^{*} $.  Remarkably that the
specific volumes of both 3-fold and 4-fold coordinated atoms do
not depend on  $z $ (or density) with  $z>z^{*} $ (5).  One
could say that $6.68{\AA}^3$ is close to a quantum limit of
3-fold coordinated carbon.  The specific volume of graphite,
$v_{3}=9{\AA}^3$ at  $T=0 $, is far above this limit.  It is
worth noticing that the calculated  $v_3 $ of a-C at is close to
specific volumes of some stable graphens with large distortions
of the $sp^2$-network.  In onions the interlayer distance for
inner layers is about $2.2\AA $ and average volume per atom is
about $6{\AA}^3$.  A similar estimation can be obtained for the
volume per atom within inner tubes of multi-wall nanotubes.
In \cite{k} the quenched glassy carbon of high density,
($3.2\ g/cm^3,z=3.8 $) in result of relaxation at $T=1500\ K$
transforms into stable a-C with $c_{3}=0.75,c_{4}=0.25 $ and  $
\rho =2.9\ g/cm^3$.  No essential changes of  $z $ and  $\rho $
were observed on the ending stage of the annealing.  Kelires has
pointed out that the revealed DLC in its nature is a type of GLC
because of low $c_4$.  The estimated specific volume,  $\bar v
$, in this a-C is  $\approx 6.75\AA^3$ what is in nice harmony with
obtained numbers in our simulations.  To investigate the
densities of states of phonons and electrons a TBMD simulation
method was used. Cells of up to $N=212$ were cut from the
prepared in CEPS samples. Afterward the TBMD simulations were
performed to provide the structure relaxation and to get the
densities of states.  It was found a considerable relaxation of
the samples on the initial stage of this procedure. The CEPS give
more crude simulation than the TBMD method (for details
see \cite{ll}). Despite changes of density and coordination
in result of the relaxation are not so large, changes of the
electron and phonon densities of states of the samples in initial
states and after relaxation are significant.

On Fig. 6 the electron
density of states for as obtained (dashed line) and TBMD-relaxed
sample (solid line) with $z<z^{*} $ and  $z>z^{*} $ are shown. It
 is seen that the CEPS gives more non-equilibrium configurations
and, as a result, a maximum of electron states at Fermi level,
$E_f $, exists.  After TBMD-relaxation a pseudogap of about
$2$ eV width forms around  $E_f $.  Very similar electron
densities of states are obtained for a-C in [11-13]
in TB and Car-Parinello approaches. A minor difference of
the electron densities of states of the structures with  $z<z^{*}
$ and $z>z^{*} $ can be recognized.  It means that the shape of
$\pi$ -electron band is not more sensitive to topological
disorder when the mobility edge is absent and no extended states
do exist.  The phonon densities of states are more sensitive to
the network local order and topology, because of the changes of
the elastic moduli.  The participation ratio (doted lines on
Fig. 7) is lower with $z>z^{*}$ for the high frequency phonons.

\begin{figure}
\begin{center}
\epsfig{figure=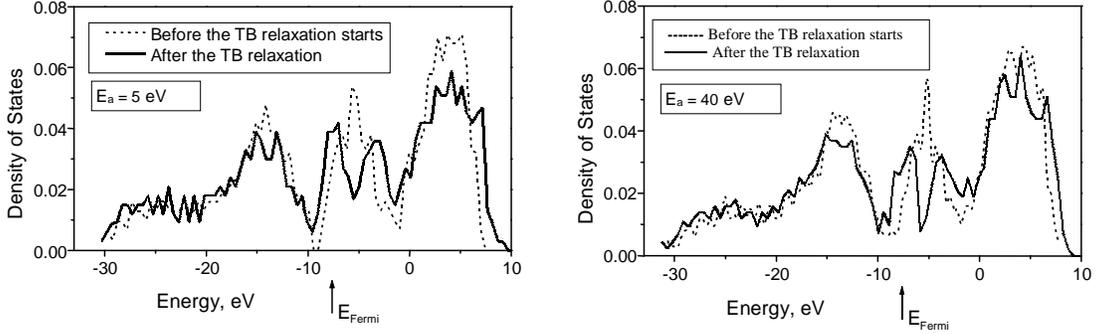,width=15cm}
\caption{\label{Fig6} Electron  densities of states, TBMD
simulations }
\vspace{2cm}
\end{center}
\end{figure}

(Because the rigidity of a-C is higher with  $z>z^{*} $, the main
maximum of the phonon density of states is shifted from
$\nu=10$ THz  at $ z<z^{*}$ to   $ \nu=30$ THz at $z>z^{*}$).
It means that the localization radius of the high frequency
phonons decreases with densification and with increasing
distortions around of $sp^2$-bonded atoms. It is worth to note
that because the size of the simulation cell is not so large
($L<10a$, $a$ is the average interatomic distance) the calculated
participation ratio here is rather a qualitative characteristic
of the localization phenomenon.  The used atomic beam deposition
procedure is based on fast quenching of as deposited atoms. The
obtained GLC ($z<z^{*}$) and DLC ($z>z^{*} $) have essentially
different structures and properties and coexist at the same  $P,T
$ just because the time of the depositing atom thermalization and
$t_{obs}$ are shorter than $\tau_{SRO}$ (\ref{ff}).
Nevertheless the revealed topological transition in the vicinity
of $z^{*}$ is a good starting point to look for metastable
polymorphous states of a-C.  For this purpose formation and
evolution of a-C under fulfilled condition (\ref{dd}) have
to be performed (see below).

\begin{figure}
\begin{center}
\epsfig{figure=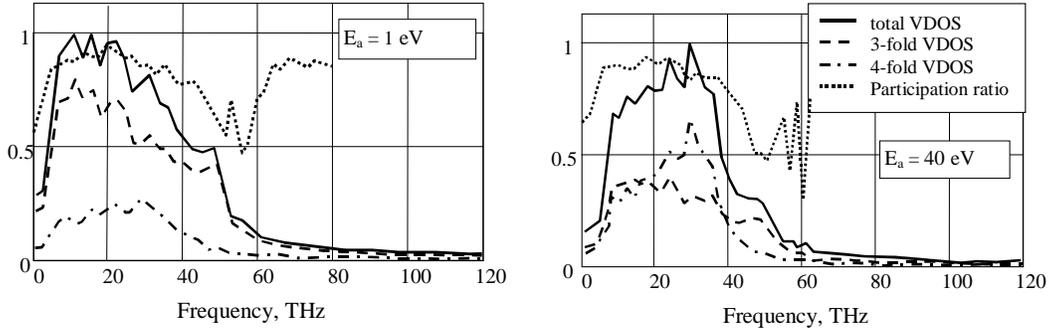,width=14cm}
\caption{\label{Fig7} Vibrational densities of states, TBMD simulations}
\end{center}
\end{figure}

\section{Phase states of a-C}
As a result of annealing at
$T\geq 1000\ K$ and $P=0$ samples of a-C
deposited from the atom beam,
independently of $z$ in initial state,
transforms into state with $ z\approx 3.21$
and $\rho\approx 2.61\ g/cm^3$. In these
simulations (CEPS) we have used cells
with $ N=776$ and $N=516$. In the initial states the
samples had  $z=3.3<z^{*},\ \rho=2.62\ g/cm^3$,
and  $z=3.4>z^{*},\ \rho=3.2\ g/cm^3$.
The annealing times were up to $15$ ps. No significant
changes of states were recognized at the end of the annealing
procedure. The obtained by the annealing metastable state of a-C
has properties close to that of GLC ($ z<z^{*}$)
formed by beam deposition. The obtained result shows that at
$P=0$, GLC is the equilibrium phase of a-C.

Because in
crystalline states along with the stable low pressure phase,
graphite, exists also diamond which is stable at high pressure,
one could expect that a-C also possesses a phase state which is
stable at high pressures. To check this idea we have investigated
the states of a-C stabilized at $0\leq P\leq 200 $ GPa and
$10^{3}\leq T\leq 7\cdot 10^{3}$ K.
\begin{figure}
\begin{center}
\epsfig{figure=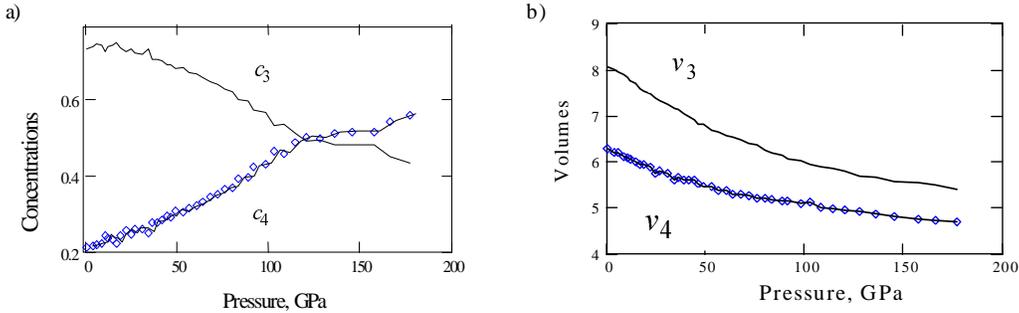,width=14cm}
\caption{\label{Fig8} Graphs of (a) concentrations $c_3,\ c_4$,
and (b) volumes $v_3,\ v_4$ as functions of $P$ at $T=2000$ K.}
\end{center}
\end{figure}

Two sets of runs were performed.
In the first of them the temperature
was kept constant. In the second set
the pressure was constant.
The simulations at constant temperature
were performed as following. Prepared by atom beam deposition sample
\\($z=3.28,\rho=2.64\ g/cm^3$) after
relaxation at a chosen temperature
($1000,\ 1500,\  2000,\ \\ 2500,\ 3000$ K) was compressed step by step.
At each of the steps the linear size of the cell was reduced on
$0.05\AA$.  Then the sample was equilibrated in thermostat during
$6$ ps before the next compression step started. One run time
was up to $314$ ps.  A similar procedure was applied in
simulations with constant pressure. At all pressures\\ ($P=1.5,\ 20,\
50,\ 100,\ 150 $ GPa) initial temperature was chosen to be equal to
$2500$ K. It increased in successive steps with step width
$\Delta T=100$ K.

As an example, specific volume, average volumes $v_3$ and $v_4$, and $z$ vs.
$P$ at $T=2000$ K are shown on Fig. 8. The pressure dependence of the
specific volume is essentially nonlinear (Fig. 8a) due to changes of LO
and coordination. Interestingly that compressibilities of $v_3$ and $v_4$
both are equal to about $10^{-3}$ GPa at $P>100$ GPa what is almost 4 times
smaller than those at $P<10$ GPa. It means that with $z\approx 3.5$ the
environments of 3-fold coordinated sites are as rigid as those of the
4-fold coordinated sites. In other words, not coordination but topology
of the carbon CRN determines macroscopic properties of a-C.

Results of simulations at constant $P$ are
depicted on Fig. 9. Specific volume vs. $T$ is shown on Fig. 9a.
\begin{figure}
\begin{center}
\epsfig{figure=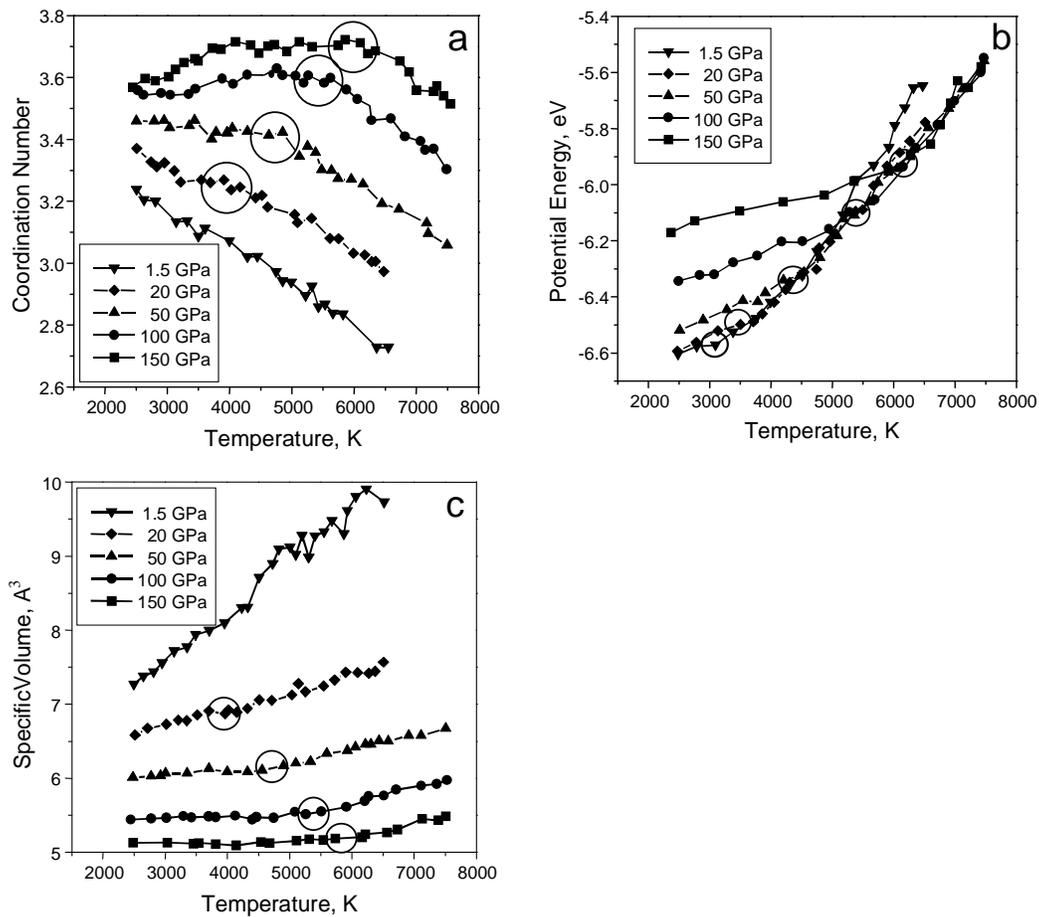,width=14cm}
\caption{\label{Fig9} (a) Coordination number, (b) potential energy,
and (c) specific volume of a-C
as functions of temperature at fixed
pressure.  Temperature intervals where slopes of the curves are
changing are marked by circles}
\end{center}
\end{figure}
It is seen that the
compressibility changes dramatically at $P>50\ $ GPa, as it took
place also at lower temperatures. Besides, the thermal expansion
coefficient (it is proportional to slope of  $v(t)$) and the
heat capacity (slope of the potential energy,  $E_{p}$, vs
$T$) change considerably in narrow temperature regions which
are marked on Fig. 9 by circles.  It is naturally to treat the
marked temperatures as  $T_{g}(P)$ because not first but
second derivatives of the free energy change step-like at the
glass transition. The slope  $dT_{g}/dP$ equals about
$20$ K/GPa. As it is known,  $T_g$ depends on the heating
and compression rates.  In our simulations these rates were
varying not so much to recognize this dependence.  The results of
the investigation of the a-C phase states are summarized on
10 where the levels of constant  $z$ on the ($P,T$) plane
are shown.
Because the pair correlation lengths are not
considerably dependent on ($P,T$), the equation
\begin{equation}
z(P,T)=const
\end{equation}
determines curves of isoconfigurational states.
\begin{figure}
\begin{center}
\epsfig{figure=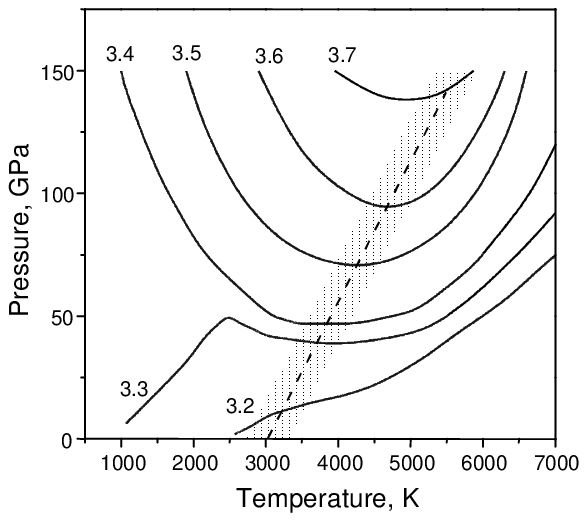,width=9cm}
\caption{\label{Fig10} The map of isoconfigurational states
(see text).  The region of glass-to-liquid transition is
shadowed.}
\end{center}
\end{figure}
\begin{figure}
\begin{center}
\epsfig{figure=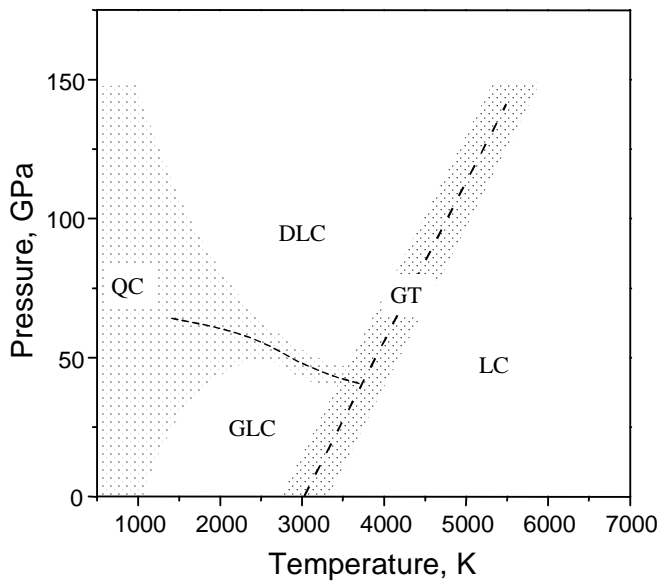,width=9cm}
\caption{\label{Fig11} Phase diagram of a-C regions
(CEPS results) of the stability, DLC, quenched a-C (QC) and liquid
C (LC) are shown. The region of GLC-to-DLC transition and the
region of glass-to-liquid transition (GT) are shadowed}
\end{center}
\end{figure}

The performed
investigations show that the states with  $z\geq 3.4$
have structure and physical properties considerably
different from those of states with $z\leq 3.3$.
Topologies of the isoconfigurational state curves with $z\geq 3.4$ and
$z\leq 3.3$ are also essentially different. Therefore we can
conclude that a~-C possesses two phase states, GLC and DLC in the
used here terminology. The phase diagram of a-C is shown on
Fig. 11. The transition region between GLC and DLC is dashed. The
glass-to-liquid transition region is also shown. The applied
approach do not allow to conclude whether first order phase
transition GLC-to-DLC exists or not. It
worth noting that continuous phase transformation without phase
transition, are also possible, especially in amorphous states
[14-17]. This kind of the phase transformation
takes place when heterophase fluctuations are strong in the
vicinity of the coexisting curve. To solve the problem of the
phase transitions of a-~C additional investigations have to be
done.  As it was pointed above, the structure relaxation kinetics
is too slow to be essential below $T=1000$ K. Therefore on the
map (10.) this temperature region belongs to the quenched a-C
states, (QC), in which appreciable short range reordering is not
seen.

\section{Concluding remarks}
Results of the simulations surprisingly support an old idea on the mechanism
of DLC formation under ion beam deposition. In accordance with this
idea energetic atom or ion forms small (its size depends on the atom energy)
region in which local temperature and pressure are rather high to initiate
$sp^3$ binding formation. If the pressure increases with the
energy, $E_a$, then the concentration
$ c_4$ has also to increase.
Experimental observations agree with idea. In [18-21]
the empiric expressions are proposed and used to
estimate  $c_{4}(E_{a})$.  Evidently the idea of the fast
equipartition of kinetic energy in a compact region on the
ballistic stage, when a few atoms are involved in collisions, is
incorrect. Just on this stage majority of inelastic replacements
and displacements, including bonding changes and formation,
happen.  Afterward some structure relaxation takes place on the
short stage of thermalization and quenching. Nevertheless it
turns out that in total complicate collisional and thermalization
kinetics of the a-C formation by beam deposition looks like
result of generation of statistically independent temperature and
pressure microspikes. Let us assume that the structure relaxes
toward equilibrium (for the current  $P $
and  $T$) state and then this last one occurs quenched.
With this assumption it has to be concluded that $P<50$ GPa at
$E_{a}=10$ eV, and $P>50$ GPa with $E_{a}=40$ eV
with effective temperature $3000\ldots 5000$ K. In \cite{fv}
effective pressure was estimated to be about $10$ GPa in result
of an empiric expression fitting. The collision dynamics of
depositing atoms bonding rearrangements was investigated in a
density functional based TBMD scheme in \cite{uf,ufl}. It
was found that below $E_{a}=30$ eV atoms
surface rearrangements dominates. The fraction of $sp^3$ bonds
considerably increases at  $E_{a}\approx 30\ eV $ when subsurface
processes dominate. These results are in qualitative agreement
with basic ideas formulated in [18-21], but no
molten states were identified in the simulations.

Another
evident idea concerns rigidity of a-C.  Bulk modulus of graphite
is almost twice smaller than that of diamond. For this reason it
could be expected that the rigidity of a-C increases rapidly with
$c_4$ above the site percolation threshold which is equal to
$c_{4}^{*}\approx 0.43$ for diamond-like lattice. In other
words, GLC-to-DLC topologic transition takes place at
$z=z^{*}\approx 3.4$ \cite{s}.  Moreover the bulk modulus of
DLC has to increase as
\begin{equation} \label{gam}
(z-z^{*})^{\gamma}+const,\ \ \gamma=3.3
\end{equation}
above the percolation
threshold (see \cite{kw,be,bg}). The elastic moduli of a-C
above  $z^{*}$ increases very fast (see Fig.2) but the law
(\ref{gam}) evidently is out of rule because, as it was shown
above,  $v_3$ changes rapidly and rigidities of the $sp^2$- and
$sp^3$-networks in the vicinity $ z^{*}$ are almost equal each to
other.

Nevertheless
manifestations of the percolation transition of $sp^3$-network are
considerable in the temperature dependencies of  $z $  at
$P>50\ GPa$. As it is seen from Fig. 8 and Fig. 9,  $z(T) $  is flat or
has a positive slope below  $T_{g}(P) $   for the states with
$z>3.4$, i.e. with  $z>z^{*} $.   This
result can be interpreted as following.  Above the percolation
threshold the $sp^3$-network forms a rigid skeleton which in
average have smaller thermal expansion coefficient than that of
the $sp^2$-network.  Therefore fringes of the $sp^3$-network play
role of compression cell walls for $sp^2$ bonded atoms. When
temperature increases $sp^2 \rightarrow
sp^3$ transformations take place due to thermoelastic stresses.
Above  $T_{g}(P) $  the skeleton is molten
and, as result, average coordination drops.  The basic problem of
any simulations is to ascertain to which extension a model mimics
reality. The same is with used here approaches.  Tersoff's
version of the potential energy allows correctly simulate
graphite and diamond structure at low temperatures. But it is not
so good (as it is pointed out i.~e.  in \cite{gr,gre}) for
simulations at big  $T $, in the vicinity of  $T_m $  and
above  $T_m $. For example, in Tersoff's version of CEPS the
diamond melting temperature is equal to $6200$ K \cite{ter}
which is about $1700$ K larger than the experimentally
determined quantity.  Currently we are investigating a-C states
taking into account the torsion energy. Boundaries of DLC and GLC
occupation regions have to be changed,  $T_g$
-curve has to be shifted, but schematically the phase state
map of a-C is expected to have a similar scheme. Account of the
torsion energy impacts on equilibrium concentrations
$c_{2},c_{3},c_{4}$   and, consequently, on topology of the
isoconfigurational curves.  Determined in \cite{gr}
$T_m$ of diamond is close to the experimentally measured
quantity. It is expected that $T_{g}(P)$
will be lower than that established in the reported here
simulations.  This expectation is based on the empirical rule:
\begin{equation}
T_{g}\approx 2T_{m}/3
\end{equation}
which valid for all glass-forming substances. For
example, with  $T_{m}=6200$ K it has to be
$T_{g}=4600$ K what is in accord with presented above
results.

If topology and connectivity of the $sp^3$-network plays a
decisive role in GLC-to-DLC transition then the curve
\begin{equation}\label{gg}
c_{4}(P,T)=c_{4}^{*}
\end{equation}

determines the boundary between GLC and DLC phases on the
map of isoconfigurational states. Here  $c_{4}^{*} $ is the
percolation threshold of the $sp^3$-network. The equation
(\ref{gg}) determines the curve of topologic phase
transition.  The site probability to belong to the percolation
cluster is its order parameter. The topologic phase transition in
a-C occurs directly connected with thermodynamic phase transition
since  $c_4 $ depends on $P,T $
and dynamic, and thermodynamic quantities are sensitive to
the CRN topology changes in the vicinity of
 $c_{4}^{*}$.  If an amorphous phase with  $ c_4 $
exist then phase transition between this phase and a phase
with  $c_{4}\not= 0 $ is unavoidable. It is the first order
phase transition if  $c_4 $  changes step-like across the
coexistence curve.  Otherwise it is the phase transition of a
higher order.

As it was reported in \cite{gre} $sp^2$-liquid (the liquid
with overwhelming 3-fold coordination of atoms) does not exist
above the melting point of graphite as it follows from CEPS of
carbon in which the torsion energy is taken into account (it is
because the torsion energy diminishes entropy of $sp^2$ bonded
structures as compare to $sp$- and $sp^3$-bonding).  At lower
temperatures role of the entropy in structure formation becomes
less important. With that $sp^2$-liquid can appear above  $T_g $
of GLC. In this case $sp^2$-liquid-to-$sp$-liquid transition
exists.  Currently hunting of this transition is in progress.

\end{document}